\documentclass[11pt, onecolumn, draftclsnofoot, peerreview]{IEEEtran}
\ifCLASSINFOpdf
\else
\fi

\usepackage[dvips]{graphicx}
\usepackage{psfrag}
\usepackage[cmex10]{amsmath}
\usepackage{amsfonts,amssymb}
\usepackage{mathrsfs}
\newtheorem{proposition}{Proposition}
\newtheorem{corollary}{Corollary}
\newtheorem{lemma}{Lemma}

\newtheorem{definition}{Definition}
\newcommand{\sps}{\scriptsize}

\newcommand{\e}[1]{\mathbb{E}\left\{#1\right\}}

\begin{document}
%
\title{On Distribution Preserving Quantization}
%
%
%

\author{Minyue~Li, Janusz~Klejsa, and~W.~Bastiaan~Kleijn \thanks{M.~Li, and J.~Klejsa are with the School of Electrical Engineering, KTH Royal Institute of Technology, Stockholm, Sweden. e-mail: \{minyue.li, janusz.klejsa\}@ee.kth.se} \thanks{W.~B.~Kleijn is both with the School of Engineering and Computer Science, Victoria University of Wellington, Wellington, New Zealand and with the School of Electrical Engineering, KTH Royal Institute of Technology, Stockholm, Sweden. email: bastiaan.kleijn@ecs.vuw.ac.nz}}

\maketitle

\begin{abstract}
Upon compressing perceptually relevant signals, conventional quantization generally results in unnatural outcomes at low rates. We propose distribution preserving quantization (DPQ) to solve this problem. DPQ is a new quantization concept that confines the probability space of the reconstruction to be identical to that of the source. A distinctive feature of DPQ is that it facilitates a seamless transition between signal synthesis and quantization. A theoretical analysis of DPQ leads to a distribution preserving rate-distortion function (DP-RDF), which serves as a lower bound on the rate of any DPQ scheme, under a constraint on distortion. In general situations, the DP-RDF approaches the classic rate-distortion function for the same source and distortion measure, in the limit of an increasing rate. A practical DPQ scheme based on a multivariate transformation is also proposed. This scheme asymptotically achieves the DP-RDF for i.i.d.~Gaussian sources and the mean squared error.
\end{abstract}


%
\IEEEpeerreviewmaketitle

\section{Introduction}
\IEEEPARstart{Q}{uantization} is an integral component in the lossy coding of perceptually relevant signals, e.g., audio and video. Conventional quantization seeks the optimal trade-off between the rate and a (perceptual) distortion measured on a pair of realizations in the sample space. Such a paradigm allows the reconstruction to converge to the source as the rate increases, thus achieving the best possible quality. However, at certain low rates, merely optimizing the rate against the distortion, compared to other strategies, e.g., synthesizing the signal from a model, can lead to \emph{unnatural} reconstruction. Examples include the following facts that are widely known about practical quantizers:
\begin{itemize}
  \item
  The reconstruction is discrete-valued. In image coding, the discrete nature of the reconstruction severely affects the rendering quality. A popular remedy is halftoning \cite{Ulichney1987}, which attempts to transform discretized images into continuous-tone images;
  \item
  The reconstruction has limited bandwidth. This causes a so-called ``band-limited'' artifact in audio compression \cite{Liu2008}. Bandwidth extension (BWE) \cite{Jax2006} and spectral band replication (SBR) \cite{Dietz2002} have been developed to solve this problem. They are essentially based on synthesizing the missing frequency bands;
  \item
  The reconstruction is reduced to zero at a rate of zero, even if a probabilistic model of the source is known to the decoder. In fact, if a model of the source is available, a reconstruction can be synthesized. Synthesis, although it may not be optimal in the rate-distortion sense, can produce \emph{natural} reconstruction. Analysis-synthesis is a common substitute of quantization for low rate coding of perceptually relevant signals (see, e.g., \cite{McAulay1986, Aizawa1995}).
\end{itemize}

It can be seen that a premise of compressing perceptually relevant signals is the \emph{naturalness} of the reconstruction. Generally, a signal is judged as being natural if it has a high occurrence probability in nature. This notion is in line with the widespread belief that the neural processing is adapted to the environment \cite{Simoncelli2001}. We note that the naturalness can be related to the context, e.g., to judge the naturalness of a speech signal, one may consider only how it is compared to natural speech signals. Natural signals can be modeled as a probability space. Quantization is a system that operates on this probability space, resulting in another that describes its reconstruction. To achieve a general naturalness of the reconstruction, it is logical to restrict the two probability spaces to be close.


The probability space of natural signals is required to be known \emph{a priori}. Studies on the statistics of natural signals, e.g., natural sound \cite{Bell1996} and images \cite{Simoncelli2001}, can help in obtaining such a probability space. From an information theoretical perspective, the probability space of the source, which is assumed to be revealed to both the encoder and the decoder, can be utilized. Supposing that the source is natural, it forms a subspace of the probability space of natural signals. Thus preserving the source probability space can fulfill our goal of ensuring the naturalness of the reconstruction. An advantage of adopting the source probability space is that no additional definitions are needed than those used in classic rate-distortion theory.

With conventional quantization, the probability space of the reconstruction generally differs from that of the source. This deviation is not only an implementational limitation but is often a theoretical necessity. Quantization has its roots in rate-distortion theory \cite{Berger1971}, which, among many things, defines the \emph{rate-distortion function} (RDF) that provides the optimal achievable rate-distortion trade-off for any quantizer. In general situations, the reconstruction that achieves the RDF forms a different probability space from the source. This is reflected by the following facts:
\begin{itemize}
\item
The optimal reconstruction is discrete-valued for many sources at a certain squared error distortion \cite{Rose1994};
\item
The source is the sum of the reconstruction and a quantization noise that is independent of the reconstruction, when the Shannon lower bound is tight;
\item
When the rate is zero, the reconstruction becomes the mathematical expectation of the source for the minimum mean squared error.
\end{itemize}
These results are consistent with the aforementioned facts of practical quantizers, implying that the problem of conventional quantization in altering the source probability space needs to be solved on a theoretical level. Since statistical differences are, \emph{per se}, measures on probability measures, it can be difficult, if at all possible, to achieve the preservation of the source statistics in classic rate-distortion theory by choosing a sample-based distortion measure. An alternative approach is to impose constraints on the probability measure of the reconstruction.
%

An early attempt of introducing constrains on the reconstructed statistics to quantization is \emph{moment preserving quantization} \cite{Delp1991}. The preservation of certain statistical moments turned out to be advantageous in the context of image coding \cite{Delp1979}. However, moment preserving quantization has some limitations: 1) it cannot preserve non-moment-like statistical properties, e.g., the continuous range of the sample values; 2) it is based on arranging space partition and reconstruction points, so is limited by the rate of the quantizer; and 3) its performance depends on a specific choice of the moments to be preserved and is, therefore, difficult to analyze.

In this article, we consider a new class of quantization, namely \emph{distribution preserving quantization} (DPQ). DPQ preserves the \emph{probability space} of the source, by which \emph{all} statistical properties are maintained. Instead of manipulating the parameters of any particular quantizer, DPQ uses an ensemble of quantizers, which, as a whole, achieves the preservation of the source probability space. Such a construction facilitates DPQ with no restriction on the rate. Moreover, the preservation of the probability space facilitates analysis.

DPQ provides a link between conventional quantization and synthesis. In the zero-rate situation (omitting model description), the reconstruction has to be generated in the same manner as the source. When the rate is higher, less synthesis is needed and DPQ can become more like conventional quantization, which at high rates already preserves the probability space of the source \emph{in some senses}. A key feature of DPQ is that it can achieve a seamless transition from one technique to the other. In particular, this is a natural outcome of optimizing a rate-distortion trade-off on top of the preservation of probability space.



The authors of this article have proposed practical DPQ schemes in \cite{Li2009, Li2010}, which have shown a superior performance over conventional quantizers in audio coding. This article is dedicated to some theoretical aspects of DPQ. In particular, we study an amended rate-distortion theory that serves as a guideline of DPQ. The main contributions of this article can be summarized as follows:
\begin{itemize}
  \item a formal definition of DPQ (Section \ref{sec:dpq-definition});
  \item a lower bound of the rate-distortion performance of DPQ schemes, namely the \emph{distribution preserving rate-distortion function} (DP-RDF), and its properties (Section \ref{sec:DP-RDF});
  \item an asymptotically optimal DPQ scheme and its properties (Section \ref{sec:transf-dpq}); and
  \item a proof of the achievability of the DP-RDF for Gaussian distribution and the mean squared error (Section \ref{sec:achievability}).
\end{itemize}

%

\section{Definition of DPQ}
\label{sec:dpq-definition}
With conventional quantization, the reconstruction can form an arbitrary probability space, which does not guarantee a perceived naturalness.
A solution is to confine the probability space of the reconstruction to be identical to that of the source, which is the essence of DPQ. An alternative is to relax the identity by putting a constraint on a measure of two probability spaces. One example can be found in \cite{Li2009}, where a Kullback-Leibler divergence is used as such a measure. However, we impose the two probability spaces to be equivalent in this article. A benefit is that no more mathematical entities are needed than those used in classic rate-distortion theory.

The probability space of the source and the reconstruction are denoted as $(A, \mathscr{A}, \mu)$ and $(\tilde{A}, \tilde{\mathscr{A}}, \tilde{\mu})$, respectively. Here $A$ is a \emph{sample space} consisting of all realizations of the source, $\mathscr{A}$ is a \emph{$\sigma$-algebra} consisting of subsets of $A$, and $\mu$ is a \emph{probability measure} on $\mathscr{A}$;  $\tilde{A}$, $\tilde{\mathscr{A}}$, and $\tilde{\mu}$ are defined similarly. Quantization can be described as a mathematical structure that links the two probability spaces.

Conventional quantization is defined as a mapping from $A$ to $\tilde{A}$. If DPQ is also defined as a mapping, it must be a \emph{measure-preserving transformation}. However, a measure-preserving transformation does not facilitate data compression, since the entropy is invariant. To obtain a feasible definition for DPQ, stochastic codes must be introduced. According to Billingsley \cite{Billingsley1965}, a \emph{stochastic code} is a channel, in which the key component is a \emph{conditional probability measure} $\phi$ defined on the Cartesian product of $A$ and $\tilde{\mathscr{A}}$, denoted as $A \times \tilde{\mathscr{A}}$. The conditional probability measure $\phi$, together with the probability measure of the source $\mu$, induces a source-reconstruction joint probability space $(A \times \tilde{A}, \mathscr{A} \times \tilde{\mathscr{A}}, \rho)$, where
\begin{align}
  \rho(G, \tilde{G}) = \int_G \phi(a, \tilde{G}) d \mu(a), \label{eqn:joint_measure}
\end{align}
for any $G \in \mathscr{A}$ and $\tilde{G} \in \tilde{\mathscr{A}}$. It further determines the probability measure of the reconstruction, i.e.~$\tilde{\mu}$, as
\begin{align}
  \tilde{\mu}(\tilde{G}) = \rho(A, \tilde{G}). \label{eqn:reconstr_measure}
\end{align}
By choosing a proper $\phi$, it is possible to achieve $\tilde{\mu}(G) = \mu(G), \forall G \in \mathscr{A}$, thus fulfilling the requirement of DPQ.

Although most of the existing quantization methods are deterministic, stochastic codes do exist in practice. An example is the dithered quantization \cite{Gray1993}, for which a dither is generated by a random number generator, added to the source and subtracted from the output of a quantizer, yielding a final reconstruction. With a proper dither, the dithered quantization is statistically equivalent to a channel with an additive noise \cite{Schuchman1964}, i.e., $\phi(a, \tilde{G}) = \varepsilon\{n: a + n \in \tilde{G}\}$, where $\varepsilon$ is the probability measure of the additive noise.

However, defining a DPQ as a stochastic code is not constructive, i.e., it does not lead to a practical encoder and decoder. A better definition resorts to a quantizer ensemble. A \emph{quantizer ensemble} is a probability space $(Q, \mathscr{Q}, \psi)$, where $Q$ consists of measurable mappings from $A$ to a countable subset of $\tilde{A}$, $\mathscr{Q}$ denotes a $\sigma$-algebra of subsets of $Q$, and $\psi$ is a probability measure on $\mathscr{Q}$. A quantizer ensemble is independent of the source. It can be seen that $Q$ consists of quantizers within the classic definition. For each use of the quantizer ensemble, a quantizer in $Q$ is randomly selected according to $\psi$ to perform the quantization. We note that a conventional quantizer is a special quantizer ensemble.

A quantizer ensemble is a stochastic code, incurring a conditional probability measure:
\begin{align}
\phi(a, \tilde{G}) = \psi\{q: q(a) \in \tilde{G}\}, \label{eqn:ensemble2stoch}
\end{align}
for any $a \in A$ and $\tilde{G} \in \tilde{\mathscr{A}}$. The probability measure of the reconstruction of a quantizer ensemble can then be described by (\ref{eqn:joint_measure}) and (\ref{eqn:reconstr_measure}). The randomness of a quantizer ensemble facilitates flexibility of the statistical properties of the reconstruction.

\begin{figure}
\centering
\psfrag{S}[r][r]{\sps Source}
\psfrag{C}[l][l]{\sps Reconstruction}
\psfrag{E}[c][c]{\sps Encoder}
\psfrag{D}[c][c]{\sps Decoder}
\psfrag{Q}[c][c]{\sps Quantizer}
\psfrag{O}[c][c]{\sps Ensemble}
\psfrag{R}[c][c]{\sps Randomness}
\includegraphics[width=0.5\columnwidth]{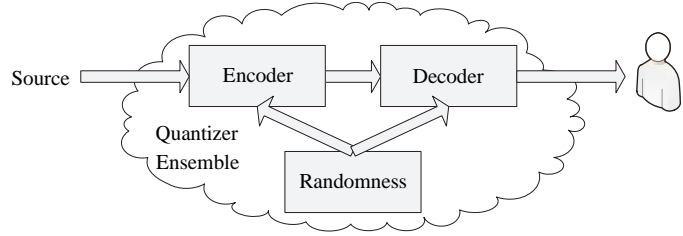}
\caption{A typical source coding scenario with a quantizer ensemble.}
\label{fig:ensemble}
\end{figure}

Although the quantization is a deterministic operation given the selected quantizer, observers outside the ensemble have no knowledge about the selection, so perceive it as operating stochastically. Figure \ref{fig:ensemble} illustrates a typical source coding scenario, where the quantizer is split into an encoder and a decoder, both of which can utilize some randomness that is unknown to the observer. A synchronicity between the encoder and the decoder, \emph{if needed}, can be achieved by using pseudo-random number generation. We note that such synchronization mechanism is not always needed. It is true that the encoder and the decoder should jointly behave as a stochastic code, but either of them can be deterministic. It is also possible that they both are stochastic, but have independent randomness.

Based on the notion of a quantizer ensemble, we can now define DPQ.
\begin{definition}[DPQ]
  Distribution preserving quantization is a quantizer ensemble, for which the probability space of the reconstruction is identical to that of the source.
\end{definition}

As mentioned, an essential consideration for DPQ is the rate-distortion trade-off, similar as conventional quantization. The rate and the distortion are both well defined for conventional quantizers and hence for the elements of a quantizer ensemble. We denote $\bar{D}: Q\rightarrow [0, \infty)$ and $\bar{R}: Q \rightarrow [0, \infty)$ as the distortion and the rate of an individual quantizer in a quantizer ensemble. For a particular quantizer $q \in Q$, the distortion $\bar{D}$ is the expectation of a distortion measure $e: A \times \tilde{A} \rightarrow [0, \infty)$, i.e.,
\begin{align}
  \bar{D}(q) = \int_A e(a, q(a)) d\mu(a).
\end{align}
All individual quantizers in a quantizer ensemble share the same distortion measure, and the distortion of the quantizer ensemble for that distortion measure is defined as the expected distortion of an individual quantizer, i.e.,
\begin{align}
  D &= \int_Q \bar{D}(q) d\psi(q) \nonumber\\
  &= \int_A \int_Q e(a, q(a)) d\psi(q) d\mu(a) \nonumber\\
  &= \int_A \int_{\tilde{A}} e(a, b) d\phi(a, b) d\mu(a) \nonumber\\
  &= \int_{A \times \tilde{A}} e(a, b) d\rho(a, b).
\end{align}
In addition, each individual quantizer is associated with a uniquely decodable code. The expected codeword length defines its rate. The rate of the quantizer ensemble is defined as the expected rate of an individual quantizer, i.e.,
\begin{align}
  R &= \int_Q \bar{R}(q) d\psi(q).
\end{align}

In the following, we give two examples of DPQ.

\subsection{Simple Example of DPQ}
\label{sec:simple}
A trivial DPQ scheme is to generate reconstructions according to the probability measure of the source but statistically independently of the source. To describe it formally, we let $Q$ be all single-image mappings: $Q = \{q: q(a) = q(b), \forall a,b \in A\}$ and
\begin{align}
  \psi\{q: q(a) \in \tilde{G}\} = \mu(\tilde{G}), \label{eqn:simple_dpq}
\end{align}
for any $a \in A$ and $\tilde{G} \in \tilde{\mathscr{A}}$. Applying (\ref{eqn:joint_measure}), (\ref{eqn:ensemble2stoch}) and (\ref{eqn:simple_dpq}), We can verify the identity of the source and the reconstruction in terms of the probability structure and the independence between them by
\begin{align}
  \rho(G, \tilde{G}) &= \int_G \psi\{q: q(a) \in \tilde{G} \} d \mu(a) = \int_G \mu(\tilde{G}) d \mu(a) = \mu(G)\mu(\tilde{G}).
\end{align}

Obviously, the rate of this DPQ scheme can be made to zero. An interesting fact of this scheme is that it describes the principle of synthesis-based reconstruction. Provided a source model, such a method generates reconstructions without the need of additional information about any particular realization of the source. The model may involve some transmission but is assumed in the context of rate-distortion theory as \emph{a priori} knowledge about the source probability space.

A pitfall of this simple DPQ scheme is that it leads to a fixed distortion. It is desirable that the distortion decreases when the rate for describing a particular source realization increases.

\subsection{DPQ Derived from Any Quantizer}
Another implementation of DPQ can be obtained by extending any conventional quantizer $q_0$, which maps $A$ to countable subset $\tilde{A}_0 \subseteq \tilde{A}$, to a quantizer ensemble and assigning a proper measure to it. Specifically, given an output of $q_0$, the quantizer reconstructs the source by randomly sampling among the values that can result in the same output, according to the relative probabilities of these values. A special case of this methodology can be found in \cite{Li2009}. In the language of the quantizer ensemble, this DPQ consists of $Q = \{q: q_0(q(a)) = q_0(a), \forall a \in A\}$. Defining $q_0^{-1}(\tilde{a}) = \{a: q_0(a) = \tilde{a}, a \in A\}$ for $\tilde{a} \in \tilde{A}_0$, we can write the probability measure of the quantizer ensemble as
\begin{align}
  & \psi\{q: q(a) \in \tilde{G}\} = \left\{ \begin{array}{ll}
  \frac{\mu\{q_0^{-1}(q_0(a)) \cap \tilde{G}\}}{\mu\{q_0^{-1}(q_0(a))\}} & \mu\{q_0^{-1}(q_0(a))\} \neq 0 \\
  0 & \mu\{q_0^{-1}(q_0(a))\} = 0
  \end{array} \right. ,
\end{align}
for any $a \in A$ and $\tilde{G} \in \tilde{\mathscr{A}}$. We verify the probability space preservation of this scheme by showing
\begin{align}
  \tilde{\mu}(\tilde{G}) &= \rho(A, \tilde{G}) = \int_A \psi\{q: q(a) \in \tilde{G} \} d \mu(a) \nonumber\\
  &= \sum_{\tilde{a} \in \tilde{A}_0, \mu\{q_0^{-1}(\tilde{a})\} \neq 0} \int_{q_0^{-1}(\tilde{a})} \frac{\mu\{q_0^{-1}(\tilde{a}) \cap \tilde{G}\}}{\mu\{q_0^{-1}(\tilde{a})\}} d \mu(a) \nonumber \\
  &= \sum_{\tilde{a} \in \tilde{A}_0, \mu\{q_0^{-1}(\tilde{a})\} \neq 0} \mu\{q_0^{-1}(\tilde{a}) \cap \tilde{G}\} \nonumber\\
  &= \mu\{\tilde{G}\}.
\end{align}

With this DPQ scheme, one can compromise between the rate and the distortion. However, from heuristics we may find that the distortion of this scheme can be relatively large. In terms of the mean squared error (MSE), this DPQ loses $3$ dB against $q_0$ \cite{Li2009} at the same rate.

In \cite{Li2010}, a DPQ that achieves a better performance was proposed. An obvious question is: what is the optimal trade-off between the rate and the distortion for DPQ and how to achieve it? A large part of this article is devoted to answer these questions. Before moving to this discussion, we define the scope of this article.

\subsection{Scope of This Article}
The definition of DPQ given earlier is based on the probability space, which makes it suitable for sources with an abstract alphabet. It is possible to follow such a notion for further discussion of DPQ, similarly as the treatment of \cite{Dobrusin1963} and \cite[Chapter 7]{Berger1971}. However, in the context of practical quantization, the sample space of the source mostly refers to the $k$-dimensional Euclidean space $\mathbb{R}^k$ with some $k$, and the source is then known as a random vector $X$, which consists of $k$ random variables (r.v.). The probability measure of such a probability space can be fully described by a probability distribution function of the random vector, $F_X$, which is also known as the cumulative distribution function (c.d.f.). Confined to the language of random variables, we can define a quantizer ensemble as a bivariate function $\tilde{X} = q(X | \Theta)$, where $\tilde{X}$ denotes the reconstruction and $\Theta$ is an auxiliary random vector that governs the selection of a quantizer for a use of the ensemble. The $\Theta$ is independent of $X$. The stochastic code incurred by such a quantizer ensemble can be described by a conditional probability distribution function $F_{\tilde{X}|X}$.

On quantizing a sequence of random vectors, DPQ needs to preserve the joint probability distribution of the entire sequence. This article mainly deals with the DPQ for sequences that are comprised of independently and identically distributed (i.i.d.) random vectors. For such a source, DPQ aims to preserve the marginal probability distribution of each random vector and the independence among the random vectors. We note that, if the marginal probability distribution of a random vector in the sequence is preserved by one use of a quantizer ensemble, and the uses of the quantizer ensemble on different random vectors are independent, the quantizer ensemble is a DPQ for the entire sequence.

\section{Distribution Preserving Rate-Distortion Function}
\label{sec:DP-RDF}
The RDF plays an important role in lossy source coding. It gives a guideline of the minimum rate that any quantizer can achieve, subject to a constraint on the distortion between the source and its reconstruction. Here we define a similar function for DPQ. The function is referred to as the distribution preserving rate-distortion function (DP-RDF). It serves as a lower bound of the achievable rate of any DPQ scheme under a constrained distortion.

\begin{definition}[DP-RDF]
The distribution preserving rate-distortion function for probability distribution $F_X$ and a distortion measure $e$ is defined as
\begin{align}
R_{\mathrm{DP}}(D) = \inf_{ F_{\tilde{X}|X} \in \mathcal{F}(D)} I(X; \tilde{X}),
\end{align}
where
\begin{align}
  \mathcal{F}(D) = \left\{F_{\tilde{X}|X}: \mathbb{E}\{e(\tilde{X}, X)\} \leq D, F_{\tilde{X}}(x) = F_X(x), \forall x \right\}.
\end{align}
\end{definition}

Next we show that the DP-RDF is a lower bound for DPQ. To get there, we first show that
\begin{lemma}
  The DP-RDF is a non-increasing convex function.
\end{lemma}

\begin{IEEEproof}
  To prove that the DP-RDF is non-increasing, we consider any $0 \leq D_2 \leq D_1$. It follows that $\mathcal{F}(D_2) \subseteq \mathcal{F}(D_1)$ and therefore, $R_{\mathrm{DP}}(D_2) \geq R_{\mathrm{DP}}(D_1)$.

  To prove the convexity, we assume that conditional probability $F_1$ achieves $R_{\mathrm{DP}}(D_1)$ and $F_2$ achieves $R_{\mathrm{DP}}(D_2)$.  For any $0 \leq \lambda \leq 1$, let $D = \lambda D_1 + (1 - \lambda) D_2$. We consider a conditional probability $F = \lambda F_1 + (1 - \lambda) F_2$. It can be seen that $F \in \mathcal{F}(D)$, so with $X$ and $\tilde{X}$ induced by $F$, $R_{\mathrm{DP}}(D) \leq I(X ; \tilde{X})$. Since the mutual information is a convex function of a conditional probability function (see, e.g., \cite[Theorem 2.7.4]{Cover2006}), we also find that $I(X ; \tilde{X}) \leq \lambda R_{\mathrm{DP}}(D_1) + (1 - \lambda) R_{\mathrm{DP}}(D_2)$. So the DP-RDF is a convex function.
\end{IEEEproof}

Then we show that DP-RDF serves as a lower bound for DPQ in the following proposition.
\begin{proposition}[DP-RDF is a lower bound for DPQ]
Consider $k$ i.i.d.~random vectors $X_1, \cdots, X_k$ (denoted as $X_K$), each of which follows probability distribution $F_X$. Given any DPQ scheme $\tilde{X}_K = q(X_K| \Theta)$, we consider a \emph{single-letter fidelity criterion} $e_k$, which is derived from a distortion measure $e$ as
\begin{align}
  e_k(x_K, \tilde{x}_K) = k^{-1} \sum_{i=1}^k e(x_i, \tilde{x}_i).
\end{align}
If the expected fidelity of the DPQ satisfies $\mathbb{E}\{e_k(X, \tilde{X})\} \leq D$, the per-dimension rate of the DPQ must be greater than or equal to the DP-RDF for $F_X$ and $e$.
\end{proposition}

\begin{IEEEproof}
The proof is analogous to the proof of the converse of source coding theorem (see, e.g., \cite[Chapter 10.4]{Cover2006}). The critical steps of the proof are the following:
\begin{align}
  k^{-1} R &= k^{-1} \mathbb{E}\{\bar{R}(\Theta)\} \nonumber\\
  & \geq k^{-1} H(\tilde{X}_K|\Theta) \label{eqn:DP_rdf_low_1}\\
  & = k^{-1} \left( H(\tilde{X}_K|\Theta) - H(\tilde{X}_K | X_K , \Theta) \right) \nonumber\\
  & = k^{-1} \left( h(X_K|\Theta) - h(X_K | \tilde{X}_K, \Theta) \right) \nonumber\\
  & \geq k^{-1} \left( h(X_K) - h(X_K | \tilde{X}_K) \right) \nonumber\\
  & = k^{-1} I(X_K; \tilde{X}_K) \nonumber\\
  & \geq  k^{-1} \sum_{i=1}^k I(X_i; \tilde{X}_i) \nonumber \\
  & \geq  k^{-1} \sum_{i=1}^k R_{\mathrm{DP}} \left( \e{e(X_i, \tilde{X}_i)} \right) \label{eqn:DP_rdf_low_4}\\
  & \geq  R_{\mathrm{DP}} \left( k^{-1} \sum_{i=1}^k \e{e(X_i, \tilde{X}_i)} \right) \label{eqn:DP_rdf_low_5}\\
  & = R_{\mathrm{DP}}\left( \e{e_k( X_K, \tilde{X}_K )} \right) \nonumber\\
  & \geq R_{\mathrm{DP}}(D), \label{eqn:DP_rdf_low_6}
\end{align}
where (\ref{eqn:DP_rdf_low_1}) is due to the fact that the rate of each individual quantizer in a quantizer ensemble is greater than or equal to the entropy of its reconstruction; (\ref{eqn:DP_rdf_low_4}) holds because DPQ preserves the joint probability distribution of $X_K$ and hence must preserve the marginal probability distribution of each random vector, so $I(X_i; \tilde{X}_i)$ must be bounded by the DP-RDF; (\ref{eqn:DP_rdf_low_5}) is based on the convexity of the DP-RDF; and (\ref{eqn:DP_rdf_low_6}) exploits the monotonicity of the DP-RDF.
\end{IEEEproof}

For a general source and distortion measure, whether the DP-RDF can be achieved by a DPQ scheme is an open problem. However, we will show in this paper that the DP-RDF for a Gaussian distribution and MSE is achievable.

In the remainder of this section, we derive the DP-RDF for Gaussian distributions and MSE, then compare this DP-RDF to the corresponding RDF. We also try to link the DP-RDF to the RDF for general sources and distortion measures.

\subsection{DP-RDF for Gaussian Distributions and MSE}
Similarly to the RDF, it is generally difficult to obtain the DP-RDF analytically. However, the Gaussian source with MSE is one of the cases that the DP-RDF is derivable. In lossy source coding, such a source and distortion measure is usually of particular interest.
\begin{proposition}
  \label{pro:Gaussian_rd}
  The DP-RDF for a Gaussian distribution and MSE is
  \begin{align}
  R_{\mathrm{DP}}(D) = \left\{ \begin{array}{ll}
    \log \frac{\sigma_X^2}{\left(\sigma_X^2 D - D^2/4\right)^{\frac{1}{2}}} & D < 2 \sigma_X^2 \\
    0 & D \geq 2 \sigma_X^2
  \end{array} \right. ,
  \label{eqn:gaussian_rd}
  \end{align}
  where $\sigma_X^2$ represents the variance of the Gaussian distribution.
\end{proposition}

\begin{IEEEproof}
Let $\tilde{X}$ be an r.v.~that follows the same probability distribution as $X$. Since the mean and the variance of $\tilde{X}$ equal those of $X$, we find
\begin{align}
  D &= \mathbb{E}\left\{(X - \tilde{X})^2\right\} \nonumber \\
  &= \mathbb{E}\left\{(X - \mu_X)^2\right\} + \mathbb{E}\left\{(\tilde{X} - \mu_X)^2\right\} - 2\mathbb{E}\left\{(X - \mu_X) (\tilde{X} - \mu_X)\right\} \nonumber\\
  &= 2 \sigma_X^2 - 2 \mathbb{E}\left\{(X - \mu_X) (\tilde{X} - \mu_X)\right\},
\end{align}
where $\mu_X$ denotes the mean of $X$. Therefore the covariance matrix of the joint random vector $[X, \tilde{X}]^{\mathrm{T}}$ is
\begin{align}
  C = \begin{bmatrix}
    \sigma_X^2 & \sigma_X^2 - D/2\\
    \sigma_X^2 - D/2 & \sigma_X^2
  \end{bmatrix}.
\end{align}
This matrix is positive semi-definite if and only if $D \leq 2 \sigma_X^2$. Knowing the covariance matrix, the differential entropy of a random vector is upper bounded \cite[Theorem 8.6.5]{Cover2006}. Using this property and the fact that the differential entropy of $\tilde{X}$ equals that of $X$, we obtain
\begin{align}
  I(X; \tilde{X}) &= h(X) + h(\tilde{X}) - h(X, \tilde{X}) \nonumber\\
  &\geq \log\left(2\pi \mathrm{e} \sigma_X^2 \right)  - \log\left(2\pi \mathrm{e} \det(C)^{\frac{1}{2}} \right) \nonumber \\
  &= \log \sigma_X^2 - \log \left(\sigma_X^2 D - D^2/4\right)^{\frac{1}{2}}.
\end{align}
The equality is achieved when $[X, \tilde{X}]^{\mathrm{T}}$ is jointly Gaussian distributed. It is easy to show that this condition is fulfilled without violating the preservation of the source probability distribution. We hence have verified (\ref{eqn:gaussian_rd}) for $D \leq 2\sigma_X^2$. Then using the non-increasing property of the DP-RDF, we can verify it for $D > 2\sigma_X^2$.
\end{IEEEproof}

\begin{figure}
\centering
\psfrag{R}[c][c]{\sps Rate (nat)}
\psfrag{M}[c][c]{\sps MSE}
\psfrag{F}[l][l]{\sps RDF}
\psfrag{D                        }[l][l]{\sps DP-RDF}
\psfrag{1}[lc][lc]{\sps $0.01$}
\psfrag{2}[c][c]{\sps $0.1$}
\psfrag{3}[c][c]{\sps $1$}
\psfrag{4}[c][c]{\sps $2$}
\psfrag{5}[c][c]{\sps $5$}
\psfrag{6}[c][c]{\sps $0$}
\psfrag{7}[c][c]{\sps $1$}
\psfrag{8}[c][c]{\sps $2$}
\includegraphics[width=0.5\columnwidth]{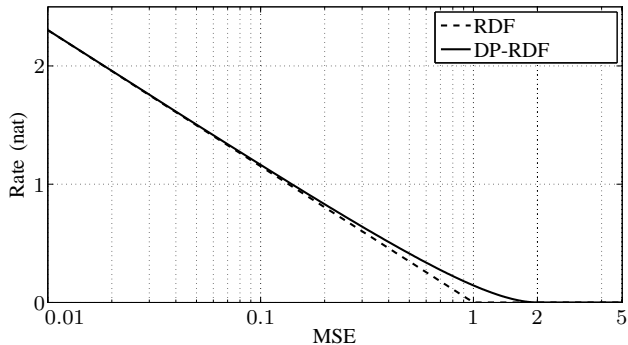}
\caption{The RDF and the DP-RDF for the standard Gaussian distribution and MSE.}
\label{fig:rdf_dpcrd}
\end{figure}

We compare the DP-RDF to the RDF for the same Gaussian distribution and MSE, which is
\begin{align}
  R(D) = \left\{ \begin{array}{ll}
    \frac{1}{2} \log \frac{\sigma_X^2}{D} & D < \sigma_X^2 \\
    0 & D \geq \sigma_X^2
  \end{array}\right. .
\label{eqn:uni_gaussian_DP_rd}
\end{align}
Figure \ref{fig:rdf_dpcrd} illustrates both functions. It can be seen that the minimum MSE of the DP-RDF is $2 \sigma_X^2$ when the rate is zero. This is an achievable rate-distortion pair for DPQ, and the simple DPQ scheme in Section \ref{sec:simple} naturally achieves it. For the RDF, the minimum MSE at zero-rate is $\sigma_X^2$, half of that in the case of the DP-RDF. The gap between the DP-RDF and the RDF is simply because DPQ randomly generates a reconstruction according to the source probability distribution, while an MSE-optimized quantizer outputs the mean of the source, when the rate is zero. In general, the requirement of probability distribution preservation increases the distortion. This loss, however, can vanish at high rates. For a Gaussian distribution and MSE, we see that
\begin{align}
  \lim_{D \rightarrow 0} R_{\mathrm{DP}}(D) - R(D)
  = \lim_{D \rightarrow 0} \frac{1}{2} \log \frac{\sigma_X^2}{\sigma_X^2 - \frac{D}{4}}
  = 0.
\end{align}
The behavior of the DP-RDF at low rates and high rates implies that the optimal DPQ forms a transition between synthesis and conventional quantization.

Proposition \ref{pro:Gaussian_rd} also leads to a conceptually optimal construction of DPQ for a Gaussian r.v.~and MSE, which is given by the following corollary.
\begin{corollary}
\label{cor:awgn}
For a Gaussian r.v.~$X$ with mean $\mu_X$ and variance $\sigma_X^2$, consider another Gaussian r.v.~$N$ that is independent of $X$ and has zero-mean and variance $\sigma_N^2$. The following r.v.,
\begin{align}
  \tilde{X} = \left(\frac{\sigma_X^2}{\sigma_X^2 + \sigma_N^2}\right)^{\frac{1}{2}} (X - \mu_X + N) + \mu_X,
\end{align}
has the same probability distribution as $X$ and the mutual information between $X$ and $\tilde{X}$ achieves the DP-RDF for $X$ and the MSE between $X$ and $\tilde{X}$.
\end{corollary}

\begin{IEEEproof}
First, given the fact that $N$ is independent of $X$ and is Gaussian distributed with zero-mean and variance $\sigma_N^2$, it is clear that $\tilde{X}$ follows the same Gaussian distribution as $X$. Then with simple algebra, one can obtain the MSE between $X$ and $\tilde{X}$ as
\begin{align}
  D = \mathbb{E}\left\{(X - \tilde{X})^2 \right\} = 2 \sigma_X^2 \left( 1 - \left( \frac{\sigma_X^2}{\sigma_X^2 + \sigma_N^2} \right) ^ {\frac{1}{2}} \right).
\end{align}
Finally, the mutual information between $X$ and $\tilde{X}$ can be written as
\begin{align}
  R &= I(X; \tilde{X}) = I(X; X + N) = \frac{1}{2} \log \frac{\sigma_X^2 + \sigma_N^2}{\sigma_N^2} = \log \frac{\sigma_X^2}{\left(\sigma_X^2 D - D^2/4\right)^{\frac{1}{2}}}. \label{eqn:cor_1}
\end{align}
Comparing (\ref{eqn:cor_1}) to Proposition \ref{pro:Gaussian_rd}, we verify Corollary \ref{cor:awgn}.
\end{IEEEproof}

Corollary \ref{cor:awgn} indicates that if there is a quantizer that operates like an additive white Gaussian noise (AWGN) channel and has a rate equal to the capacity of the channel, an optimal DPQ for a Gaussian r.v.~and MSE is such a quantizer followed by a shifting and a scaling. It is known that entropy coded dithered lattice quantization (ECDQ) behaves effectively as a channel with additive noise and the rate equals the channel capacity \cite{Zamir1992}. Unfortunately, when ECDQ has finite dimensionality, the quantization noise is not Gaussian. However, a DPQ scheme can be obtained by applying a \emph{non-linear} transformation after an ECDQ. This approach will be discussed later in this article.

We have compared the DP-RDF and the RDF for a Gaussian distribution and MSE. We now try to analyze their relationship for more general sources and distortion measures.

\subsection{Relationship between DP-RDF and RDF}
The relationship between the RDF and the DP-RDF is usually not as straightforward as in the case of Gaussian distributions and MSE. However, we will show that for a broad class of sources and distortion measures, the DP-RDF approaches the corresponding RDF when the rate increases.

\begin{figure}
\centering
\psfrag{X}[c][c]{\sps $X$}
\psfrag{Y}[c][c]{\sps $\bar{X}$}
\psfrag{Z}[c][c]{\sps $\tilde{X}$}
\psfrag{U}[c][c]{\sps $W$}
\psfrag{V}[c][c]{\sps $\bar{W}$}
\psfrag{+}[c][c]{\sps $+$}
\includegraphics[width=0.35\columnwidth]{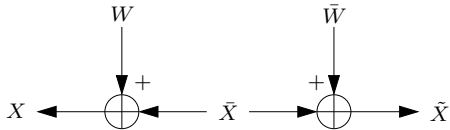}
\caption{A backward-forward channel that preserves the probability distribution of the source. In this model, $\bar{X}$ is independent of $W$ and $\bar{W}$. The probability distribution of $W$ equals that of $\bar{W}$.}
\label{fig:lsb_dpq}
\end{figure}

It is known that the RDF equals the Shannon lower bound (SLB), when the source and its reconstruction are related by a ``backward channel'' with additive noise \cite{Berger1971}. From the reconstruction that achieves the SLB, we construct a ``forward channel'' with the same noise statistics as for the ``backward channel''. Then the output of the forward channel follows the probability distribution of the source and hence defines an upper bound on the DP-RDF. This upper bound can be related to the SLB. Figure \ref{fig:lsb_dpq} illustrates the relationship of a source $X$, an SLB achieving reconstruction $\bar{X}$, a distribution preserving output $\tilde{X}$, and the noise of a backward and a forward channel, denoted by $W$ and $\bar{W}$, respectively.

The SLB is defined for a \emph{difference} distortion measure as
\begin{align}
  R_{\mathrm{SLB}}(D_W) = h(X) - \sup_{F_W: \mathbb{E}\{e(W)\} \leq D_W } h(W). \label{eqn:slb}
\end{align}
When the distortion measure satisfies $e(w) = e(-w)$, the p.d.f.~of $W$ is symmetric. Then letting $\bar{W} = -W$ suffices for the backward-forward channel in Figure \ref{fig:lsb_dpq} to preserve the source probability distribution. In this case, the mutual information between $X$ and $\tilde{X}$ follows
\begin{align}
  I(X; \tilde{X}) &= h(X) + h(\tilde{X}) - h(X, \tilde{X}) \nonumber\\
   &= 2h(X) - h(X - \tilde{X}, X) \nonumber \\
   &= 2h(X) - h(X - \tilde{X}) - h(X | X - \tilde{X}) \nonumber \\
   &= 2h(X) - h(2W) - h(X - W), \label{eqn:slb_dpq_r}
\end{align}
where (\ref{eqn:slb_dpq_r}) stems from the fact that $W$ and $X - W$ are independent. The distortion between $X$ and $\tilde{X}$ is
\begin{align}
  D = \mathbb{E}\{e(X - \tilde{X})\} = \mathbb{E}\{e(2W)\}. \label{eqn:slb_dpq_d}
\end{align}

This rate-distortion characteristic forms an upper bound on the DP-RDF. To relate it to the SLB (\ref{eqn:slb}), we need to investigate the effect of scaling the noise on the SLB. We consider the following lemma.
\begin{lemma}
\label{lem:multiple_distortion}
  If a difference distortion measure $e$ satisfies
  \begin{align}
    e(\alpha w) = c(\alpha) e(w), \quad \alpha > 0,
    \label{eqn:multiple_distortion}
  \end{align}
  the SLB for any source and $e$ satisfies
  \begin{align}
    R_{\mathrm{SLB}}(c(\alpha) D) = R_{\mathrm{SLB}}(D) - \log \alpha.
  \end{align}
\end{lemma}

\begin{IEEEproof}
This lemma can be simply proven by
 \begin{align}
  R_{\mathrm{SLB}}(D) &= h(X) - \sup_{\mathbb{E}\{e(W)\} \leq D} h(W) \nonumber\\
  &= h(X) - \sup_{\mathbb{E}\{e(\alpha W)\} \leq c(\alpha) D} h(\alpha W) + \log \alpha \nonumber \\
  &= R_{\mathrm{SLB}}(c(\alpha) D) + \log \alpha.
\end{align}
\end{IEEEproof}

Now we can provide a relationship between the SLB and the DP-RDF.
\begin{proposition}[relationship between SLB and DP-RDF]
  For a source $X$ and a difference distortion measure $e$, if
\begin{enumerate}
  \item the distortion measure satisfies $e(2w) = c e(w)$ and $e(w) = e(-w)$, $\forall w$, and
  \item the SLB is tight, and $W$ is the reconstruction error that achieves $R_{\mathrm{SLB}}(D/c)$,
\end{enumerate}
then the DP-RDF for $X$ and $e$ is bounded by
  \begin{align}
    R_{\mathrm{SLB}}(D) \leq R_{\mathrm{DP}}(D) \leq R_{\mathrm{SLB}}(D) + h(X) - h(X - W).
  \end{align}
\end{proposition}

\begin{IEEEproof}
The left inequality is trivial: the DP-RDF is larger than or equal to the corresponding RDF, which is larger than or equal to the SLB.

To prove the right inequality, we apply the upper bound of the DP-RDF given by (\ref{eqn:slb_dpq_r}) and (\ref{eqn:slb_dpq_d}).
Using Lemma \ref{lem:multiple_distortion}
we find
\begin{align}
  R_{\mathrm{DP}}(D)
   &\leq  2h(X) - h(2W) - h(X - W) \nonumber\\
   &= R_{\mathrm{SLB}}(D/c) - \log 2 + h(X) - h(X - W) \nonumber\\
   &= R_{\mathrm{SLB}}(D) + h(X) - h(X - W).
\end{align}
\end{IEEEproof}

Since the p.d.f.~of $W$ becomes narrower when the distortion approaches $0$, $h(X)$ can get closer to $h(X-W)$, then $R_{\mathrm{DP}}(D)$ may approach $R_{\mathrm{SLB}}(D)$ and hence also the RDF. A Gaussian source with MSE is an example of this situation. For rigorous conditions of $h(X) - h(X-W) \rightarrow 0$, one may refer to \cite{Linder1994}, which also proves that the SLB is asymptotically tight under mild assumptions, as the distortion decreases. This implies that the DP-RDF is asymptotically equivalent to the RDF for a large range of sources and distortion measures.

\section{Transformation-Based DPQ}
\label{sec:transf-dpq}
In \cite{Li2010}, a scalar DPQ scheme that uses dithering and a non-linear transformation was proposed. It is based on the fact that the preservation of the source probability distribution can be obtained by performing a transformation on the output of a dithered quantizer. We refer to such a DPQ paradigm as \emph{transformation-based DPQ}. Here we generalize the idea to a vector DPQ. An extensive analysis on the transformation-based DPQ will be conducted. The analysis shows that this scheme has nice rate-distortion properties. In particular, it is able to asymptotically achieve the DP-RDF for Gaussian distributions and MSE.

\subsection{Quantization Scheme}
\begin{figure}
\centering
\psfrag{X}[b][b]{\sps $X$}
\psfrag{E}[b][b]{\sps ECDQ}
\psfrag{Y}[b][b]{\sps $\hat{X}$}
\psfrag{Z}[b][b]{\sps $\tilde{X}$}
\psfrag{D}[c][c]{\sps $Z$}
\psfrag{Q}[c][c]{\sps $q_L(\cdot)$}
\psfrag{G}[c][c]{\sps $g(\cdot)$}
\psfrag{-}[c][c]{\sps $-$}
\psfrag{+}[c][c]{\sps $+$}
\includegraphics[width=0.35\columnwidth]{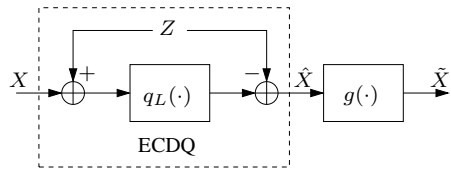}
\caption{Diagram of transformation-based DPQ.}
\label{fig:dpq}
\end{figure}


A transformation-based DPQ, as shown in Figure \ref{fig:dpq}, is a construction based on an ECDQ followed by a transformation. The rate of the ECDQ is defined as the rate of the transformation-based DPQ. ECDQ has a rate-distortion performance that is close to the RDF \cite{Zamir1992} and therefore, we can expect that the whole DPQ scheme can achieve a low distortion in a rate range where the transformation does not affect the signal significantly.

Let source $X$ be a $k$-dimensional random vector. The ECDQ uses a subtractive dither and a $k$-dimensional lattice quantizer $q_L(\cdot)$. The lattice quantizer performs the following operation:
\begin{align}
  q_L(\lambda) = l_n, \quad \lambda \in \mathcal{P}_n \nonumber ,
\end{align}
where $l_n$ and $\mathcal{P}_n$ represent the $n$-th lattice point and the $n$-th lattice cell, respectively. Every cell can be defined as a translation on a basic cell $\mathcal{P}_0$:
\begin{align}
  \mathcal{P}_n = l_n + \mathcal{P}_0 \nonumber.
\end{align}
In the following the volume of $\mathcal{P}_0$ is denoted as $V$, i.e., $V = \mathrm{Vol}(\mathcal{P}_0)$. The lattice quantizer is used together with a subtractive dither. The dither $Z$ is generated according to the uniform distribution over the basic cell $\mathcal{P}_0$. It is added to $X$ before the lattice quantization and subtracted from the quantized signal, resulting in ${\hat{X}}$. Finally, a transformation $g$ is applied to ${\hat{X}}$, yielding a reconstruction of the source block, ${\tilde{X}}$. The whole DPQ operation can be written as a bivariate function with the dither being an auxiliary variable: $\tilde{X} = g(q_L(X + Z) - Z | Z)$.

With the ECDQ, $N = \hat{X} - X$ is independent of $X$ and uniformly distributed over $-\mathcal{P}_0$ \cite{Zamir1996}. Therefore, $\hat{X}$ follows a continuous probability distribution, which can be calculated analytically.



For a one-to-one mapping $\tilde{X} = g(\hat{X})$, the p.d.f. of $\tilde{X}$ becomes
\begin{align}
  f_{\tilde{X}}(x) = f_{\hat{X}}(g^{-1}(x)) |\det (\mathfrak{J}(x))|,
\end{align}
where $\mathfrak{J}(x)$ is the Jacobian of $g^{-1}(x)$. Thus a sufficient and necessary condition for the scheme in Figure \ref{fig:dpq} to be a DPQ is
\begin{align}
  |\det (\mathfrak{J}(x))| = \frac{f_X(x)}{f_{\hat{X}}(g^{-1}(x))}, \quad \mathrm{a.e.} \label{eqn:Jocobian}
\end{align}
Except for the scalar case, it can be difficult to find a transformation that fulfills this condition. An existing method is Rosenblatt's transformation \cite{Rosenblatt1952}, which performs a sequence of transformations on a number of continuous r.v.'s to obtain independent r.v.'s that are uniformly distributed over $[0,1]$. An inverse Rosenblatt's transformation can transform independent r.v.'s that are uniformly distributed over $[0,1]$ to r.v.'s with an arbitrary probability distribution, which is known in the field of random number generation as \emph{inverse transform sampling}.

In the following, we use $X_I$ to denote a random vector that consists of a subset of the r.v.'s of another random vector $X$, where $I$ is the set that contains all chosen indices. The cardinality of $I$ is denoted as $|I|$. In addition, we use $F_{X|Y}$ and $f_{X|Y}$ to denote the conditional c.d.f.~and p.d.f.~of a random vector $X$ given another random vector $Y$.

Rosenblatt's transformation performs the following operations sequentially:
\begin{align}
  U_1 &= F_{\hat{X}_1}(\hat{X}_1) \nonumber \\
  \cdots \nonumber \\
  U_n &= F_{\hat{X}_n|\hat{X}_{\{1, \cdots, n-1\}}}(\hat{X}_n|\hat{X}_{\{1, \cdots, n-1\}}) \nonumber \\
  \cdots
\label{eqn:rosenblatt_1}
\end{align}
The result of the transformation is that $U_1, \cdots, U_k$ are independently and uniformly distributed over $[0, 1]$. We then perform an inverse Rosenblatt's transformation as follows:
\begin{align}
  \tilde{X}_1 &= F_{X_1}^{-1}(U_1) \nonumber \\
  \cdots \nonumber \\
  \tilde{X}_n &= F_{X_n|X_{\{1, \cdots, n-1\}}}^{-1}(U_n|U_{\{1, \cdots, n-1\}}) \nonumber \\
  \cdots
\label{eqn:rosenblatt_2}
\end{align}
where the inverse c.d.f.~follows a standard definition, i.e., $F_{X}^{-1}(x) = \inf\{ \tilde{x}: F_{X}(\tilde{x}) \geq x \}$. In fact, for the inverse transformation (\ref{eqn:rosenblatt_2}), reordering $U_1, \cdots, U_k$ does not influence the probability distribution of ${\tilde{X}}$. However, we consider this particular order, since it yields a small Euclidean distance between ${\hat{X}}$ and ${\tilde{X}}$. It will be shown that DPQ with this transformation leads asymptotically to the optimal rate-distortion performance in certain circumstances.

Summing up (\ref{eqn:rosenblatt_1}) and (\ref{eqn:rosenblatt_2}) we may write an overall transformation for the proposed DPQ scheme:
\begin{align}
  g_i({\hat{x}}) &= F_{X_i|X_{\{1, \cdots, i-1\}}}^{-1}\left(F_{\hat{X}_i|\hat{X}_{\{1, \cdots, i-1\}}}\left(\hat{x}_i|\hat{x}_{\{1, \cdots, i-1\}}\right) \big| g_1({\hat{x}}),\cdots,g_{i-1}({\hat{x}})\right), \quad i=1, \cdots, k.
  \label{eqn:transform}
\end{align}
From the properties of Rosenblatt's transformation, we can see that the probability distribution of the transformation output equals that of the source. This can also be verified by checking that (\ref{eqn:transform}) fulfills (\ref{eqn:Jocobian}). Therefore we can claim:
\begin{proposition}
The proposed scheme in Figure \ref{fig:dpq} with the transformation defined by (\ref{eqn:transform}) is a DPQ.
\end{proposition}

Here we briefly consider the behavior of the transformation-based DPQ at low rates and high rates, respectively. When the rate is low, the output of the ECDQ has a near-uniform probability distribution, which is reformed by the transformation to a desired shape. At high rates, the output of the dithered quantization has a probability distribution that resembles that of the source. Then the transformation modifies the ECDQ output only slightly. We will show later that, as the rate increases, the modification becomes so small that it does not increase the distortion that is introduced by the ECDQ.

In the following, we will make an extensive analysis on transformation-based DPQ. The analysis deals with its general properties and asymptotic properties w.r.t.~high rates and high dimensionality, respectively.

\subsection{Properties of Transformation-Based DPQ}
We are interested in the amount of modification that the transformation (\ref{eqn:transform}) introduces. If $g(\hat{x})$ always falls in a vicinity of $\hat{x}$, the transformation-based DPQ will have a similar rate-distortion performance as ECDQ.

The transformation (\ref{eqn:transform}) is non-linear and seems difficult to analyze. However, the transformation has a special structure, i.e., it consists of an inner and an outer function that are closely related. Therefore, some properties exist, which facilitate an analysis on the transformation-based DPQ.

We first show a property of ECDQ using the following lemma.
\begin{lemma}
  \label{lem:map}
  For a $k$-dimensional ECDQ with input $X$ and output $\hat{X}$, given any realization $\hat{x}$ of $\hat{X}$, and any two disjoint subsets $I$ and $J$ of $\{1, \cdots, k\}$, there exists an
  \begin{align}
    \tilde{x} \in \hat{x} + \mathcal{P}_0 \nonumber,
  \end{align}
  such that
  \begin{align}
    F_{X_I|X_J}(\tilde{x}_I|\tilde{x}_J) = F_{\hat{X}_I|\hat{X}_J}(\hat{x}_I|\hat{x}_J) \nonumber.
  \end{align}
\end{lemma}

\begin{IEEEproof}
  In ECDQ, the quantization noise, $\hat{X} - X$, is independent of the input and uniformly distributed over $-\mathcal{P}_0$. So the p.d.f.~of $\hat{X}$ is
  \begin{align}
    f_{\hat{X}}(\hat{x}) = \frac{1}{V} \int_{-\mathcal{P}_0} f_{X}(\hat{x} - \tau) d \tau.
  \end{align}
  Let $K = \{1,\cdots,k\}$, the marginal p.d.f.~of $\hat{X}_J$ is
  \begin{align}
  f_{\hat{X}_J}(\hat{x}_J) &= \int_{\mathbb{R}^{k - |J|}} f_{\hat{X}_{J, K\setminus J}}(\hat{x}_J, \upsilon) d\upsilon \nonumber\\
  &= \frac{1}{V} \int_{\mathbb{R}^{k - |J|}} \int_{\mathcal{P}_0} f_{\hat{X}_{J, K\setminus J}}(\hat{x}_J + \tau_J, \upsilon + \tau_{K\setminus J}) d \tau d\upsilon \nonumber \\
  &= \frac{1}{V} \int_{\mathcal{P}_0} \int_{\mathbb{R}^{k - |J|}} f_{\hat{X}_{J, K\setminus J}}(\hat{x}_J + \tau_J, \upsilon + \tau_{K\setminus J}) d\upsilon d\tau \nonumber \\
  &= \frac{1}{V} \int_{\mathcal{P}_0} f_{X_J}(\hat{x}_J + \tau_J) d\tau.
  \end{align}
  Then the conditional p.d.f.~of $\hat{X}_I$ given $\hat{X}_J$ writes
  \begin{align}
  f_{\hat{X}_I|\hat{X}_J}(\hat{x}_I|\hat{x}_J) &= \frac{f_{\hat{X}_{I, J}}(\hat{x}_{I, J})} {f_{\hat{X}_J}(\hat{x}_J)} \nonumber\\
  &= \frac{\int_{\mathcal{P}_0} f_{X_{I, J}}(\hat{x}_I + \tau_I, \hat{x}_J + \tau_J) d\tau}{\int_{\mathcal{P}_0} f_{X_J}(\hat{x}_J + \tau_J) d\tau}.
  \end{align}
  The conditional c.d.f.~can be derived as
  \begin{align}
  F_{\hat{X}_I|\hat{X}_J}(\hat{x}_I|\hat{x}_J) &= \int_{-\infty}^{\hat{x}_I}  f_{\hat{X}_I|\hat{X}_J}(\upsilon|\hat{x}_J) d \upsilon\nonumber \\
  &= \int_{-\infty}^{\hat{x}_I} \frac{\int_{\mathcal{P}_0} f_{X_{I,J}}(\upsilon + \tau_I, \hat{x}_J + \tau_J) d\tau}{\int_{\mathcal{P}_0} f_{X_J}(\hat{x}_J + \tau_J) d\tau} d \upsilon \nonumber \\
  &= \frac{\int_{\mathcal{P}_0} \int_{-\infty}^{\hat{x}_I} f_{X_{I,J}}(\upsilon + \tau_I, \hat{x}_J + \tau_J) d \upsilon d\tau}{\int_{\mathcal{P}_0} f_{X_J}(\hat{x}_J + \tau_J) d\tau} \nonumber\\
  &= \frac{\int_{\mathcal{P}_0} F_{X_I|X_J}(\hat{x}_I + \tau_I|\hat{x}_J + \tau_J) f_{X_J}(\hat{x}_J + \tau_J) d\tau}{\int_{\mathcal{P}_0} f_{X_J}(\hat{x}_J + \tau_J) d\tau}.
  \label{eqn:cond_cdf}
  \end{align}
  Because $ F_{X_I|X_J}$ is a continuous function and $f_{X_J}$ is nonnegative, Lemma \ref{lem:map} follows from the mean value theorem of integration.
\end{IEEEproof}

Lemma \ref{lem:map} implies that, for any $i$-th step of the transformation (\ref{eqn:transform}), if one is free to choose $g_1(\hat{x}), \cdots, g_{i-1}(\hat{x})$, the result of the transformation is almost surely bounded in the $\mathcal{P}_0$ vicinity of $\hat{x}$. Unfortunately, due to the sequential nature of the transformation, $g_1(\hat{x}), \cdots, g_{i-1}(\hat{x})$ are fixed for the $i$-th step of (\ref{eqn:transform}), thus there is no guarantee of a bound on the result of the transformation. However, when the source is composed of independent r.v.'s, the influence of the sequential treatment is less severe. We find the following proposition.

\begin{proposition}
\label{pro:rect}
 For a $k$-dimensional ECDQ with input $X$ and output $\hat{X}$, if $X$ is composed of independent r.v.'s, then given any realization $\hat{x}$ of $\hat{X}$, there exists an
\begin{align}
  \tilde{x} \in \hat{x} + \mathcal{T}(\mathcal{P}_0) \nonumber,
\end{align}
with $\mathcal{T}(\mathcal{P}_0)$ defined as a box that covers the basic quantization cell:
\begin{align}
  \mathcal{T}(\mathcal{P}_0) = \left\{\upsilon: \inf_{\tau \in \mathcal{P}_0} \tau_i \leq \upsilon_i \leq \sup_{\tau \in \mathcal{P}_0} \tau_i \right\} \nonumber,
\end{align}
such that
\begin{align}
  F_{X_i}(\tilde{x}_i) = F_{\hat{X}_i|\hat{X}_{\{1,\cdots ,i-1\}}}\left(\hat{x}_i|\hat{x}_{\{1,\cdots ,i-1\}}\right) \nonumber
\end{align}
holds for $i = 1,2, \cdots, k$, simultaneously.
\end{proposition}

\begin{IEEEproof}
According to Lemma \ref{lem:map}, for any $\hat{x}$ and $i$, there is an
\begin{align}
  \tilde{x}^{(i)} \in \hat{x} + \mathcal{P}_0,
  \label{eqn:vicinity}
\end{align}
such that
\begin{align}
  F_{X_i}(\tilde{x}_i^{(i)}) &= F_{X_i|X_{\{1,\cdots ,i-1\}}}\left(\tilde{x}_i^{(i)}|\tilde{x}_{\{1,\cdots ,i-1\}}^{(i)}\right) \nonumber\\
  &= F_{\hat{X}_i |\hat{X}_{\{1,\cdots ,i-1\}}}\left(\hat{x}_i|\hat{x}_{\{1,\cdots ,i-1\}}\right).
  \label{eqn:col_indep_cdf}
\end{align}
It is easy to see that
\begin{align}
  \inf_{\tau \in \mathcal{P}_0} \tau_i \leq \tilde{x}_i^{(i)} - \hat{x}_i \leq \sup_{\tau \in \mathcal{P}_0} \tau_i.
\end{align}
We take
\begin{align}
  \tilde{x} = \left(\tilde{x}_1^{(1)}, \cdots, \tilde{x}_k^{(k)}\right),
\end{align}
which proves Proposition \ref{pro:rect}.
\end{IEEEproof}

Proposition \ref{pro:rect} implies that, when the source is comprised of independent r.v.'s, the transformation (\ref{eqn:transform}) does not move its input far. So the transformation-based DPQ can have a comparable rate-distortion performance to the embedded ECDQ. Proposition \ref{pro:rect} also implies the robustness of transformation-based DPQ, i.e., even if the probabilistic model does not match the input data well, the reconstruction of the transformation-based DPQ can still be bounded.


However, when the dimensionality approaches infinity, Proposition \ref{pro:rect} can become less meaningful, since the covering box may become unbounded. For high dimensionality, the transformation has some additional properties that will be considered in Section \ref{sec:high-dim}.

\subsection{Asymptotic Properties w.r.t.~High Rates}
We have analyzed the transformation at any rate. The results show that, for a source of independent r.v.'s, the transformation performs a mild change to the ECDQ output. In the following, we further consider a high rate scenario, for which the MSE of transformation-based DPQ approaches that of ECDQ.
\begin{proposition}
\label{pro:high_rate}
Let $X$ be a source random vector consisting of independent r.v.'s, each of which has a c.d.f.~and an inverse c.d.f.~whose second derivatives are bounded almost everywhere. Assume the basic cell $\mathcal{P}_0$ of the lattice used in the transformation-based DPQ is symmetric w.r.t.~each of its coordinates, meaning if
\begin{align}
  (\tau_1, \cdots, \tau_k) \in \mathcal{P}_0,
\end{align}
with any $1 \leq i \leq k$, then
\begin{align}
  (\tau_1, \cdots, -\tau_i, \cdots, \tau_k) \in \mathcal{P}_0.
\end{align}
Then, the MSE of the transformation-based DPQ and that of the embedded ECDQ satisfy
\begin{align}
  \mathbb{E}\left\{\|X - \tilde{X}\|^2\right\} = \mathbb{E}\left\{\|X - \hat{X}\|^2\right\} + O(V^{\frac{3}{k}}).
\end{align}
\end{proposition}
The proof of Proposition \ref{pro:high_rate} resorts to the technique of Taylor series and is given in Appendix.

Due to Proposition \ref{pro:high_rate}, for independent r.v.'s and MSE, transformation-based DPQ performs equally efficient as ECDQ with an increasing rate. In addition, because the optimal ECDQ can asymptotically achieve the RDF for i.i.d.~Gaussian source and MSE as the rate and the dimensionality increase \cite{Zamir1992}, the transformation-based DPQ can also asymptotically reach the RDF and hence the DP-RDF. Moreover, it will be shown later that, for i.i.d.~Gaussian sources and MSE, the transformation-based DPQ can asymptotically reach the DP-RDF at any rate as the dimensionality increases. To get there, we will first investigate the behavior of the transformation at high dimensionality.

%

\subsection{Asymptotic Properties w.r.t.~High Dimensionality}
\label{sec:high-dim}
In this subsection, we consider the asymptotic behavior of the transformation when the dimensionality increases. In particular, with a certain sequence of lattices, the shape of the basic cells can approach a ball, and the transformation (\ref{eqn:transform}) can become simpler, especially when the source r.v.'s are independent.


Let $\mathcal{B}_k(r)$ denote a $k$-dimensional ball with radius $r$, and $B_k(r)$ denote its volume:
\begin{align}
  B_k(r) = \mathrm{Vol}(\mathcal{B}_k(r)) = \frac{\pi^{\frac{k}{2}} r^k}{\Gamma\left(\frac{k}{2} + 1\right)}.
\end{align}

The following lemma considers the average of a function in a ball when the ball's dimensionality is large. The lemma is inspired by Poincar\'{e}'s observation that, if a $k$-dimensional random vector follows a uniform distribution on a sphere (the surface of a ball), then any finite subset of the r.v.'s of the random vector follows an i.i.d.~Gaussian distribution, when $k\rightarrow \infty$. A proof of this can be found in \cite{McKean1973}. We here consider the average of a function in a ball, which should make no significant difference from its average on the surface of the ball, since a thin shell located at the surface of a ball takes all the volume of the ball when the dimensionality approaches infinity, which is known as \emph{sphere hardening}. The statement and proof of our lemma are different from the mentioned work and are shown below.
\begin{lemma}
\label{lem:sphere_gaussian}
For any integer set $I$ with finite cardinality, any function $f: \mathbb{R}^{|I|} \rightarrow \mathbb{R}$ and any $\eta > 0$, the following holds
\begin{align}
  & \lim_{k \rightarrow \infty} \frac{1}{B_k(k^{\frac{1}{2}}\eta)} \int_{\mathcal{B}_k(k^{\frac{1}{2}}\eta)} f(\tau_I) d\tau = \int_{\mathbb{R}^{|I|}} f(\tau) (2\pi\eta^2)^{-\frac{|I|}{2}} \exp \frac{-\|\tau\|^2}{2 \eta^2} d\tau.
\end{align}
\end{lemma}

\begin{IEEEproof}
Because the intersection of a ball with a hyper-plane is a ball of lower dimensionality, it can be shown that
\begin{align}
& \frac{1}{B_k(k^{\frac{1}{2}}\eta)}\int_{\mathcal{B}_k(k^{\frac{1}{2}}\eta)} f(\tau_I) d\tau = \int_{\|\tau_I\|^2 \leq k \eta^2} f(\tau_I) \frac{B_{k-|I|} \left((k \eta^2 - \|\tau_I\|^2)^{\frac{1}{2}} \right)}{B_k(k^{\frac{1}{2}}\eta)} d\tau_I.
\end{align}
Further, we find
\begin{align}
  & \frac{B_{k-|I|} \left((k \eta^2 - \|\tau_I\|^2)^{\frac{1}{2}} \right)}{B_k(k^{\frac{1}{2}}\eta)} \nonumber \\
&= \frac{\pi^{\frac{k-|I|}{2}} (k \eta^2 - \|\tau_I\|^2)^{\frac{k-|I|}{2}}} {\Gamma\left(\frac{k-|I|}{2} + 1\right)} \frac{\Gamma\left(\frac{k}{2} + 1\right)}{\pi^{\frac{k}{2}}(k^{\frac{1}{2}} \eta)^{k}} \nonumber\\
&=(\pi \eta^2)^{-\frac{|I|}{2}} \left(1 - \frac{\|\tau_I\|^2}{k \eta^2} \right)^{\frac{k-|I|}{2}} \frac{\left(\frac{k}{2} + 1\right)^{\frac{|I|}{2}}}{k^{\frac{|I|}{2}}} \frac{\Gamma\left(\frac{k}{2} + 1\right) \left(\frac{k}{2} + 1\right)^{-\frac{|I|}{2}}} {\Gamma\left(\frac{k}{2} + 1 + \frac{|I|}{2}\right)}.
\end{align}
Using the property of the ratio of two Gamma functions (see, e.g., \cite[Equation 6.1.46]{Davis1964}), we can obtain
\begin{align}
  \lim_{k \rightarrow \infty} \frac{\Gamma\left(\frac{k}{2} + 1\right) \left(\frac{k}{2} + 1\right)^{-\frac{|I|}{2}}} {\Gamma\left(\frac{k}{2} + 1 + \frac{|I|}{2}\right)} = 1.
\end{align}
Also, it is easy to find that
\begin{align}
  \lim_{k \rightarrow \infty} \left(1 - \frac{\|\tau_I\|^2}{k \eta^2} \right)^{\frac{k-|I|}{2}} = \exp \frac{-\|\tau_I\|^2}{2 \eta^2}.
\end{align}
Thus Lemma \ref{lem:sphere_gaussian} is proven.
\end{IEEEproof}

Lemma \ref{lem:sphere_gaussian} indicates that the average of a function in a high dimensional ball can be calculated in a smaller space. If the lattice cells used in the transformation-based DPQ are balls, we may use Lemma \ref{lem:sphere_gaussian} to derive the transformation (\ref{eqn:transform}) for high dimensional situations, since the conditional c.d.f.~in (\ref{eqn:transform}), which is given by (\ref{eqn:cond_cdf}), is based on the average of a function over the basic cell. Unfortunately, lattice cells cannot exactly be balls. However, a sequence of lattices can approach to a ball in various senses \cite{Poltyrev1994, Zamir1996, Forney2000}. We now show that by using a \emph{sphere-bound-achieving} lattice sequence \cite{Forney2000}, the average of a bounded function in the basic lattice cell follows the same behavior as in Lemma \ref{lem:sphere_gaussian}.
\begin{lemma}
\label{lem:lattice_gaussian}
For any integer set $I$ with finite cardinality, any bounded function $f: \mathbb{R}^{|I|} \rightarrow [-M, M]$ with some $M \geq 0$ and any $\eta > 0$, there exists a sequence of lattices with increasing dimensionality such that
\begin{align}
  & \lim_{k \rightarrow \infty} \frac{1}{V_k} \int_{\mathcal{P}_0^{(k)}} f(\tau_I) d\tau = \int_{\mathbb{R}^{|I|}} f(\tau) (2\pi\eta^2)^{-\frac{|I|}{2}} \exp \frac{-\|\tau\|^2}{2 \eta^2} d\tau.
\end{align}
where $\mathcal{P}_0^{(k)}$ and $V_k$ denote the basic cell and its volume of the $k$-th lattice.
\end{lemma}

\begin{IEEEproof}
We use a sequence of lattices that is sphere-bound-achieving. In particular, the volume of the basic cell satisfies
\begin{align}
  V_k = B_k(k^{\frac{1}{2}}\eta), \label{eqn:lattice_volume}
\end{align}
and the probability that a $k$-tuple Gaussian vector, which consists of i.i.d.~Gaussian r.v.'s with zero mean and variance $\eta^2$, falls outside $\mathcal{P}_0^{(k)}$ approaches $0$, when $k \rightarrow \infty$.

We observe
\begin{align}
\frac{1}{V_k} \int_{\mathcal{P}_0^{(k)}} f(\tau_I) d\tau &= \frac{1}{B_k(k^{\frac{1}{2}}\eta)} \int_{\mathcal{B}_k(k^{\frac{1}{2}}\eta)} f(\tau_I) d\tau \nonumber\\
&\quad + \frac{1}{B_k(k^{\frac{1}{2}}\eta)} \int_{\mathcal{P}_0^{(k)} \setminus \mathcal{B}_k(k^{\frac{1}{2}}\eta)} f(\tau_I) d\tau \nonumber\\
&\quad - \frac{1}{B_k(k^{\frac{1}{2}}\eta)} \int_{\mathcal{B}_k(k^{\frac{1}{2}}\eta) \setminus \mathcal{P}_0^{(k)}} f(\tau_I) d\tau.
\end{align}
Using the fact that $f$ is bounded, we see
\begin{align}
 &\left| \frac{1}{B_k(k^{\frac{1}{2}}\eta)} \int_{\mathcal{B}_k(k^{\frac{1}{2}}\eta) \setminus \mathcal{P}_0^{(k)}} f(\tau_I) d\tau \right| \leq \frac{\mathrm{Vol}(\mathcal{B}_k(k^{\frac{1}{2}}\eta) \setminus \mathcal{P}_0^{(k)})}{B_k(k^{\frac{1}{2}}\eta)} M.
\end{align}
When $k \rightarrow \infty$, the $k$-tuple Gaussian random vector is uniformly distributed in the ball. The sphere-bounding-achieving condition implies the following \cite{Forney2000}:
\begin{align}
  \lim_{k \rightarrow \infty} \frac{\mathrm{Vol}(\mathcal{B}_k(k^{\frac{1}{2}}\eta) \setminus \mathcal{P}_0^{(k)})}{B_k(k^{\frac{1}{2}}\eta)} = 0.
\end{align}
Thus
\begin{align}
  \lim_{k \rightarrow \infty} \frac{1}{B_k(k^{\frac{1}{2}}\eta)} \int_{\mathcal{B}_k(k^{\frac{1}{2}}\eta) \setminus \mathcal{P}_0^{(k)}} f(\tau_I) d\tau = 0.
\end{align}
Since $\mathcal{P}_0^{(k)}$ has the same volume as $\mathcal{B}_k(k^{\frac{1}{2}}\eta)$, it follows
\begin{align}
  \mathrm{Vol}(\mathcal{P}_0^{(k)} \setminus \mathcal{B}_k(k^{\frac{1}{2}}\eta)) = \mathrm{Vol}(\mathcal{B}_k(k^{\frac{1}{2}}\eta) \setminus \mathcal{P}_0^{(k)}),
\end{align}
and hence
\begin{align}
  \lim_{k \rightarrow \infty} \frac{1}{B_k(k^{\frac{1}{2}}\eta)} \int_{\mathcal{P}_0^{(k)} \setminus \mathcal{B}_k(k^{\frac{1}{2}}\eta)} f(\tau_I) d\tau = 0.
\end{align}
Then using Lemma \ref{lem:sphere_gaussian}, Lemma \ref{lem:lattice_gaussian} is proven.
\end{IEEEproof}

According to Lemma \ref{lem:lattice_gaussian}, we can see that the calculation of the conditional c.d.f.~of $\hat{X}$ (\ref{eqn:cond_cdf}) and therefore the transformation (\ref{eqn:transform}), in the limit, may not require an integration over the whole basic cell. In particular, when $X$ is composed of independent r.v.'s, the transformation (\ref{eqn:transform}) can be significantly simplified. Consider a sequence of transformation-based DPQs, which uses the lattice sequence defined in the proof for Lemma \ref{lem:lattice_gaussian}. Let $g_i^{(k)}(\hat{x})$ be the $i$-th step of the transformation of the $k$-th DPQ. When the source consists of independent r.v.'s, whose p.d.f.'s are bounded, for any particular $i$, it follows that
\begin{align}
 \lim_{k \rightarrow \infty} g_i^{(k)}(\hat{x}) & = F_{X_i}^{-1}\left( \lim_{k \rightarrow \infty} \frac{\int_{\mathcal{P}_0^{(k)}} F_{X_i}(\hat{x}_i + \tau_i) f_{X_I}(\hat{x}_I + \tau_I) d\tau}{\int_{\mathcal{P}_0^{(k)}} f_{X_I}(\hat{x}_I + \tau_I) d\tau} \right) \nonumber\\
 & = F_{X_i}^{-1} \left(\int_{\mathbb{R}} F_{X_i}(\hat{x}_i + \tau) (2\pi\eta^2)^{-\frac{1}{2}} \exp \frac{-\tau^2}{2 \eta^2} d\tau \right),
 \label{eqn:transform_asympt}
\end{align}
where
$I$ denotes $\{1, \cdots, i-1\}$. Eq.~(\ref{eqn:transform_asympt}) is valid only if the choice of $i$ is independent of $k$. However, in the transformation (\ref{eqn:transform}), $i$ goes from $1$ to $k$. To make the number of steps in the transformation independent of the dimensionality of the transformation-based DPQ, we may increase the dimensionality of a source vector by appending it with pseudo-random numbers. In this way, we can quantize a $k$-dimensional source vector with a DPQ of an arbitrarily large dimensionality. In this setup, the asymptotic behavior of the transformation, i.e.~(\ref{eqn:transform_asympt}), is valid for all the steps in the transformation.

\section{Achievability of DP-RDF for Gaussian Distributions and MSE}
\label{sec:achievability}
Based on the high dimensionality analysis of transformation-based DPQ, we can show that transformation-based DPQ can achieve the DP-RDF for i.i.d.~Gaussian sources and MSE, at any rate. We propose
\begin{proposition}
\label{pro:achievable}
Given a source consisting of i.i.d.~Gaussian r.v.'s, let $R_{\mathrm{DP}}(D)$ denote the DP-RDF for the Gaussian distribution and MSE. For any distortion level $D > 0$, there exists a sequence of DPQs with increasing dimensionality, such that the rate approaches $R_{\mathrm{DP}}(D)$, while the MSE approaches a level that is smaller than or equal to $D$.
\end{proposition}

\begin{IEEEproof}
Denote the mean and the variance of the Gaussian distribution as $\mu_X$ and $\sigma_X^2$. When $D \geq 2\sigma_X^2$, $R_{\mathrm{DP}}(D) = 0$, Proposition \ref{pro:achievable} for such a situation can be fulfilled by using the simple DPQ scheme described in Section \ref{sec:simple}. Therefore, we only need to consider the case that $0 < D < 2\sigma_X^2$.

Using the lattice sequence in the proof for Lemma \ref{lem:lattice_gaussian}, we obtain a sequence of transformation-based DPQs. Then according to (\ref{eqn:transform_asympt}), we can show that the transformation for $k$-th DPQ satisfies
\begin{align}
   & \lim_{k \rightarrow \infty} g_i^{(k)}(\hat{x}) \nonumber\\
&= F_{X_i}^{-1} \left( \int_{\mathbb{R}} \int_{-\infty}^{\hat{x}_i + \tau} (4 \pi^2 \sigma_X^2 \eta^2)^{-\frac{1}{2}} \exp \left( \frac{-(x - \mu_X)^2}{2 \sigma_X^2} + \frac{-\tau^2}{2 \eta^2} \right) dx d\tau \right)\nonumber\\
&= F_{X_i}^{-1} \left( \int_{-\infty}^{\hat{x}_i} (2\pi(\sigma_X^2 + \eta^2))^{-\frac{1}{2}} \exp \frac{-(x - \mu_X)^2}{2 (\eta^2 + \sigma_X^2)} dx  \right) \nonumber\\
&= F_{X_i}^{-1} \left( \int_{-\infty}^{\frac{\sigma_X}{(\sigma_X^2 + \eta^2)^{\frac{1}{2}}}( \hat{x}_i - \mu_X) + \mu_X } (2\pi\sigma_X^2)^{-\frac{1}{2}} \exp \frac{-(x - \mu_X)^2}{2 \sigma_X^2 } dx \right) \nonumber\\
&= \frac{\sigma_X}{(\sigma_X^2 + \eta^2)^{\frac{1}{2}}}(\hat{x}_i - \mu_X) + \mu_X. \label{eqn:linear_operator}
\end{align}
We notice that this result is very similar to Corollary \ref{cor:awgn}. In fact, by the following discussion, we will show that the rate-distortion performance of the transformation-based DPQ indeed approaches the DP-RDF for the said Gaussian distribution and MSE.

It has been shown in \cite{Erez2005} that a sphere-bound-achieving lattice sequence can also be good for quantization. According to \cite{Zamir1996}, for a lattice sequence that is \emph{good for quantization} and salifies the volume condition (\ref{eqn:lattice_volume}), the noise introduced by the ECDQ is white and has a power approaching $\eta^2$. Using the fact that the transformation approaches a linear operation, i.e.~(\ref{eqn:linear_operator}), we can show that the MSE of the transformation-based DPQ satisfies
\begin{align}
  \underline{D} &= \lim_{k \rightarrow \infty} D_k
  = \left(\frac{\sigma_X}{(\sigma_X^2 + \eta^2)^{\frac{1}{2}}} - 1\right)^2 \sigma_X^2 + \frac{\sigma_X^2 \eta^2}{\sigma_X^2 + \eta^2} \nonumber\\
  &= 2 \sigma_X^2 \left(1 - \frac{\sigma_X^2}{(\sigma_X^2 + \eta^2)^{\frac{1}{2}}} \right).
\end{align}
In addition, using lattice sequence that is good for quantization, the rate of the ECDQ and hence the rate of the transformation-based DPQ satisfy \cite{Zamir1996},
\begin{align}
   \underline{R} = \lim_{k \rightarrow \infty} R_k = \frac{1}{2} \log \frac{\sigma_X^2 + \eta^2}{\eta^2}.
\end{align}
Finally, through some elementary algebra, we can verify
\begin{align}
 \underline{R} = \log \frac{\sigma_X^2}{\left(\sigma_X^2 \underline{D} - \underline{D}^2/4\right)^{\frac{1}{2}}}.
\end{align}

Therefore, the rate and the MSE of the DPQ sequence approaches a point on the DP-RDF. By choosing the value of $\eta$, $\underline{D}$ can take any value in $(0, 2\sigma_X^2)$. Then Proposition \ref{pro:achievable} is proven.
\end{IEEEproof}

\section{Conclusions}
In this article, we proposed distribution preserving quantization (DPQ) as a new lossy source coding concept, which aims to achieve a good perceived quality of signal reconstruction for the entire range of rates. To this purpose, DPQ optimizes a rate-distortion trade-off under the constraint that the probability space of the source is preserved.

The minimum rate that any DPQ scheme can achieve, under a constraint on the distortion, is lower bounded by the distribution preserving rate-distortion function (DP-RDF). In general situations, the DP-RDF approaches the classic rate-distortion function on the same source and distortion measure, when the distortion decreases. This means that, at high rates, DPQ may perform as well as conventional quantization. At low rates, DPQ relies more on synthesis to reconstruct the source, thus maintaining good perceived quality. In particular, DPQ facilitates a seamless transition between signal quantization and synthesis.

We also proposed an asymptotically optimal DPQ scheme, namely transformation-based DPQ. This scheme is shown to be as efficient as a classic quantization scheme for the mean squared error (MSE), as the rate increases. For i.i.d.~Gaussian sources and MSE, transformation-based DPQ asymptotically achieves the DP-RDF as the dimensionality increases.


%
\appendix[A Proof of Proposition \ref{pro:high_rate}]
\begin{IEEEproof}
Let $x$ be a realization of the source random vector $X$, $n$ is a realization of the ECDQ noise $N$, which is uniformly distributed over $-\mathcal{P}_0$ and independent of the source.

Using the fact that the source is composed of independent r.v.'s, the transformation (\ref{eqn:transform}) becomes
\begin{align}
  & g_i(x + n) = F_{X_i}^{-1} \left( \frac{\int_{\mathcal{P}_0} F_{X_i}(x_i + n_i + \tau_i) f_{X_I}(x_I + n_I + \tau_I) d\tau} {\int_{\mathcal{P}_0} f_{X_I}(x_I + n_I + \tau_I) d\tau} \right),
\end{align}
where $I = \{1, \cdots, i-1\}$. Since we assume that the c.d.f.~for each source r.v.~has a bounded second derivative, using Taylor series, there exists a $W$ such that
\begin{align}
  &\left|F_{X_i}(x_i + n_i + \tau_i) - F_{X_i}(x_i) - f_{X_i}(x_i)(n_i + \tau_i)\right| \leq W(n_i + \tau_i)^2.
\end{align}
Then we can find
\begin{align}
  &\frac{\int_{\mathcal{P}_0} F_{X_i}(x_i + n_i + \tau_i) f_{X_I}(x_I + n_I + \tau_I) d\tau} {\int_{\mathcal{P}_0} f_{X_I}(x_I + n_I + \tau_I) d\tau} \leq F_{X_i}(x_i) + \epsilon_i,
\end{align}
where
\begin{align}
  \epsilon_i = f_{X_i}(x_i) \frac{n_i}{V^{\frac{1}{k}}} V^{\frac{1}{k}} + W \left( \frac{n_i^2}{V^{\frac{2}{k}}}  + \sup_{\tau \in \mathcal{P}_0} \frac{\tau_i^2}{V^{\frac{2}{k}}} \right) V^{\frac{2}{k}} \label{eqn:proof_pro_highrate_1}.
\end{align}
To obtain (\ref{eqn:proof_pro_highrate_1}), we exploited the condition that $\mathcal{P}_0$ is symmetric. Then we show that there exists an $M$ such that
\begin{align}
  g_i(x+n) &\leq F_{X_i}^{-1}(F_{X_i}(x_i) + \epsilon_i) \label{eqn:proof_pro_highrate_2} \\
  &\leq F_{X_i}^{-1}(F_{X_i}(x_i)) + \frac{\epsilon_i}{f_{X_i}(F_{X_i}^{-1}(F_{X_i}(x_i)))} + M \epsilon_i^2 \label{eqn:proof_pro_highrate_3} \\
  &= x_i + (f_{X_i}(x_i))^{-1}\epsilon_i + M \epsilon_i^2 \quad \mathrm{a.s.} \label{eqn:proof_pro_highrate_4}
\end{align}
where (\ref{eqn:proof_pro_highrate_2}) uses the non-decreasing property of inverse c.d.f., (\ref{eqn:proof_pro_highrate_3}) is due to Taylor seriers and the bound on the second derivative of the inverse c.d.f.~for each source r.v., and (\ref{eqn:proof_pro_highrate_4}) holds almost surely because $x_i$ is a realization of $X_i$. Similarly to the earlier derivation, we can have
\begin{align}
  g_i(x+n) \geq x_i + (f_{X_i}(x_i))^{-1}\delta_i - M \delta_i^2 \quad \mathrm{a.s.},
\end{align}
where
\begin{align}
  \delta_i =  f_{X_i}(x_i) \frac{n_i}{V^{\frac{1}{k}}} V^{\frac{1}{k}} - W \left( \frac{n_i^2}{V^{\frac{2}{k}}}  + \sup_{\tau \in \mathcal{P}_0} \frac{\tau_i^2}{V^{\frac{2}{k}}} \right) V^{\frac{2}{k}}.
\end{align}
Therefore
\begin{align}
  (x_i - g_i(x + n))^2 &\leq  \max\left\{ ((f_{X_i}(x_i))^{-1} \epsilon_i + M\epsilon_i^2)^2,((f_{X_i}(x_i))^{-1} \delta_i - M\delta_i^2)^2 \right\} \quad \mathrm{a.s.} \label{eqn:proof_pro_highrate_5}
\end{align}

Since $N$ is uniformly distributed over $-\mathcal{P}_0$, statistical moments of $N_i/V^{\frac{1}{k}}$ do not depend on $V$. In addition, $\sup_{\tau \in \mathcal{P}_0} \tau_i^2 / V^{\frac{2}{k}}$ does not depend on $V$. Therefore
\begin{align}
  \e{\|X - \tilde{X}\|^2} = \sum_{i=1}^k \e{ (X_i - g_i(X + N))^2 }
\end{align}
is bounded by a polynomial of $V^{\frac{1}{k}}$. To show Proposition \ref{pro:high_rate}, only terms of an order lower than $3$ in the polynomial are needed. The two terms in the maximization (\ref{eqn:proof_pro_highrate_5}) share the same terms with an order lower than $3$. By picking out these terms, we have
\begin{align}
  \e{\|X - \tilde{X}\|^2} &= \sum_{i=1}^k \e{ (X_i - g_i(X + N))^2 } \nonumber\\
  &\leq \sum_{i=1}^k \e{ N_i^2 + O(V^{\frac{3}{k}}) } \nonumber\\
  &= \e{\|X - \hat{X}\|^2} + O(V^{\frac{3}{k}}).
\end{align}
\end{IEEEproof}






\ifCLASSOPTIONcaptionsoff
  \newpage
\fi



\bibliographystyle{IEEEtran}
\bibliography{dpq_references}
\end{document}